\documentclass[preprint,twocolumn,superscriptaddress,10pt,%
floatfix]{revtex4-1}
\usepackage{graphicx}

\usepackage{amsmath,amssymb}
\usepackage{xcolor}

\usepackage{mathtools,microtype}
\mathtoolsset{showonlyrefs}

\let \vec \mathbf
\let\kappa\varkappa

\usepackage{microtype}
\begin{document}
\title{Buckling of elastic filaments by discrete magnetic moments}
\author{Horst-Holger Boltz} 
\email[]{horst-holger.boltz@phys.uni-goettingen.de}
\author{Stefan Klumpp}         
\affiliation{Institute for Nonlinear Dynamics, University of G\" ottingen, 
37077 G\" ottingen, Germany}
\date{\today}

\begin{abstract}We study the buckling of an idealized, semiflexible filament along whose contour magnetic moments are placed. {We give analytic expressions for the critical stiffness of the filament below which it buckles due to the magnetic compression. For this, we consider various scenarios of the attachment of the magnetic particles to the filament. One possible application for this model are the magnetosome chains of magnetotactic bacteria. An estimate of the critical bending stiffness indicates that buckling may occur within the range of biologically relevant parameters and suggests a role for the bending stiffness of the filament  to stabilize the filament against buckling, which would compromise the functional relevance of the bending stiffness of the used filament}.
\end{abstract}
%

\maketitle
\section{Introduction}
Magnetic nanoparticles are of interest for many applications as they allow for easily controllable manipulation and tuning material properties.~\cite{Tang2005,Majetich2011,Wang2011,Singh2014,kralj2015,Jiang2016} Specifically, linear structures \cite{Tang2005,Wang2011,kralj2015,Jiang2016}, sometimes referred to as magnetic nanochains, featuring a then controllable shape anisotropy  have gathered a lot of research interest. The magnetic interactions alone lead typically to only metastable linear assemblies, so an abundant strategy is to stabilize these by attaching the magnetic particles to a soft matrix or a linear filament.~\cite{Tang2005,Wang2011} The magnetic interaction between the attached particles can lead to a change in the elastic properties of the system. For an elastic filament this has been found to lead to a magnetic contribution~\cite{Vella2013,Kiani2015} to the bending rigidity of the filament, analogously to the electrostatic contribution~\cite{Barrat1993} for polyelectrolytes. Naturally occurring examples of such structures are magnetosome chains in magnetotactic bacteria \cite{Klumpp2016}. Magnetosomes~\cite{Bazylinski2004,Yan2012} are organelles comprised of small (typical length-scale $\sim 0.1\mathrm{\mu m}$) magnetic crystals formed by biomineralization {(usually ferrimagnetic, but small enough to have a permanent magnetization)} that are used for navigation with respect to the earth's magnetic field~\cite{Bazylinski2004,Lefevre2014}. Their form is essential for their function, i.e. it is important that the crystals are reliably assembled and held in an elongated manner. To this end the crystals are organized along the magnetosome filament \cite{Frankel2006,Komeili2006,Scheffel2006,Faivre2010,Murat2013} a filamentous structure build up from MamK \cite{Komeili2006,Draper2011}, an actin-like semiflexible polymer.

Magnetosome filaments provide an exemplary system, where two pillars of classical physics, magnetism and elasticity theory, are directly relevant for a biological function. Thus, not surprisingly there has been considerable research interest in the mechanics of magnetosome filaments and related systems, e.g Refs.\  
\citenum{Shcherbakov1997,Lins2004,Cebers2005,Abracado2011,Hall2013,Vella2013,Kornig2014,Kiani2015,Meyra2016}.

In this spirit we revisit a paradigm of classical mechanics, the buckling of a stiff rod, with the change of flavor that the compressive forces are magnetic. The model we study provides insight into the mechanics of magnetosome filaments and related systems. In addition, it serves as a nice illustration of this classical problem in mechanics, with some subtle twists, that could be realized in macroscopic magneto-elastic systems. 
\section{Model}

\begin{figure}
\includegraphics[width=.975\linewidth]{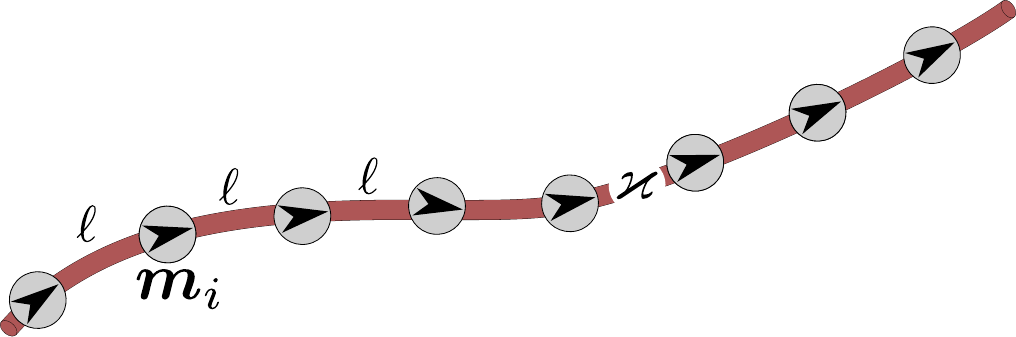}
\caption{Sketch of the system of interest: An elastic filament with finite bending rigidity $\varkappa$ with magnetic moments attached along it.}
\label{fig:sketch}
\end{figure}

A generic sketch of the system we are interested in is given by Fig.\ \ref{fig:sketch}: An incompressible rod with bending stiffness $\varkappa$ along which (point-like) magnetic {(dipole)} moments (magnetization $\vec{m}$) are placed. In general there are $N$ moments connected by $N-1$ rod sections of arc-length $\ell$, For simplicity, we only consider two moments here, that is we neglect all interactions except those between nearest neighbors as magnetic dipole-dipole interactions decay rapidly (as the cube of inverse distance). The error due to this simplification should be most notable as the filament is buckled and the spatial separation of the magnetic moments is reduced. However, the influence of higher order terms to the onset of buckling should be negligible. We will briefly discuss this later on.

One big advantage of considering only two magnetic moments is that the magnetic problem thus becomes one-dimensional (in real-space at least, with maybe one additional dimension for the relative orientation) and the filament will be oriented in this dimension and one dimension normal to it (with a rotational degeneracy), resulting in a very simple planar problem.

An important notion at this stage is that the simplification done by considering only a pair of magnetic moments is only of technical nature: only numerical details would change if one were to consider a more complex system. The conclusion that the effective bending rigidity $\alpha$, which we will define below, should be at least of order one is unaffected by this.

We introduce some nomenclature: The arc-length of the rod in between these two moments is called $\ell$ whilst their distance vector is called $\vec r$ with $\lvert \vec r\rvert =r$. We assume that the magnetic moments only exert localized forces onto the rod due to their attachment. Furthermore, we neglect the possibility of detachment meaning that the moments cannot move along the rod.

The bending energy of a (Kirchhoff) rod of length $L$ is given as the integral of the squared curvature\cite{LL7, Kirchhoff}, 
$
E_b = \frac{\varkappa}{2} \int_0^\ell \!\mathrm{d}s\, \left(\partial_s^2 \vec{q}(s)\right)^2 \text{.}
$
Here, $\vec{q}(s)$ is the space-curve that gives the shape of the central fiber of the rod parametrized in arc-length $s$, i.e. $\left(\partial_s \vec{q} (s)\right)^2 = 1$. We can write this as
$
E_b = \frac{\varkappa}{2} \int_0^\ell \!\mathrm{d}s\,  K(s)^2
$
where $K(s)$ is the local curvature.  Using the Frenet-Serret formulae\cite{docarmo1976} (i.e. working with the local tripod of tangent~$\vec t$, normal~$\vec n$ and binormal vector~$\vec b$)
{
\begin{align}
\partial_s \vec{t} &= + K \vec n &
\partial_s \vec n &= - K \vec t
\label{eq:fs}
\end{align}}%
and suitable initial conditions the curvature incorporates the total geometrical information and is sufficient to reconstruct the rod's shape.

{Throughout this work, we assume that fluctuations due to temperature are negligible. For the elastic energy used here, this is justifiable\cite{kierfeld2010} as long as one considers length-scales $\ell$ that are not large against the rod's persistence length $\ell_p = \kappa/(k_B T)$. For higher temperatures, one is essentially left with (entropic) springs connecting the magnetic moments, a system which has been studied earlier in Ref. \citenum{annunziata}. Also, we focus on the onset of buckling and, thus, limit ourselves to a regime where the elastic energy is comparable to the magnetic contributions. If this is not the case, our model turns into the well-studied\cite{cerda2013,sanchez2013} system of a polymer of magnetic colloids allowing for a different kind of buckling transition that has been observed experimentally\cite{huang2016}. Furthermore, neglecting (thermal) fluctuations {as well as torsional long-scale deformations} makes the problem strictly two-dimensional (the binormal $\vec b$ is a constant) which facilitates analytical progress but is certainly not valid deep in the buckled regime.} 

No external forces act on the section of the rod between the two magnetic moments, thus its shape has to obey the Euler-Lagrange equation associated with the bending energy. This leads to the famous problem of the Eulerian {\em elastica} curve, that is we are looking for solutions of (neglecting torsion)  \cite{Singer2008}
\begin{align}
\partial_s^2K&=\lambda K - \frac{K^3}{2} \label{eq:dgl}
\end{align}
with $\lambda$ being a Lagrange parameter associated with the incompressibility of the rod. {For the sake of self-containedness, we give a brief derivation of this equation in the appendix}.

The simplest solution of eq.\ \eqref{eq:dgl} is a circle, i.e. a planar curve of constant curvature $K_0=R^{-1}$ with $R$ being the circle's radius. It's obvious that this is indeed a solution of eq.\ \eqref{eq:dgl} for $K(0)=K_0$, $\partial_s K(0)=0$ and $\lambda=K_0^2/2$. The bending energy of the connecting filament following an arc of a circle with radius $R$ is
\begin{align}
E_b &= \frac{\varkappa \ell}{2} R^{-2} \text{.} \label{eq:ebcirc}
\end{align}
The circular solution to the elastica problem of eq.\ \eqref{eq:dgl} is the only one for which there is a simple, closed expression for the bending energy and we will therefore employ numerical methods to determine it if the circle is not the appropriate solution. Furthermore, we argue below that buckling patterns extending over multiple magnetic moments are negligible. 
The total magnetic energy of the two magnetic moments is given by the dipole-dipole interaction
\begin{align}
E_m&=-r^{-3} \frac{\mu_0}{4\pi} \left( \frac{ 3 \left(\vec{m_1}\cdot \vec{r}\right) \left( \vec{m_2} \cdot \vec{r} \right)}{r^2} - \vec{m_1}\cdot\vec{m_2} \right) \text{.}
\end{align}
{One could account for the effect of a higher number of magnetic moments (i.e. the repetition of the filament) by replacing the strength of the magnetic moments $m = \lvert\vec{m}\rvert$ with an effective rescaled one. For example, for an infinite chain of magnetic dipole moments, we would have to change $m$ according to $m^2\rightarrow 2\zeta(3) m^2 \approx 2.404 m^2$ with $\zeta(n)$ being the Riemann zeta function. However, this would make our analysis explicitly dependent on the total number of magnetic moments and, thus, include another degree of freedom. As stated before, this wil only lead to a different numerical prefactor, so we focus on the case of only two magnetic moments.
}

\section{Free alignment} 
\label{sec:free}

\begin{figure}
\includegraphics[width=.975\linewidth]{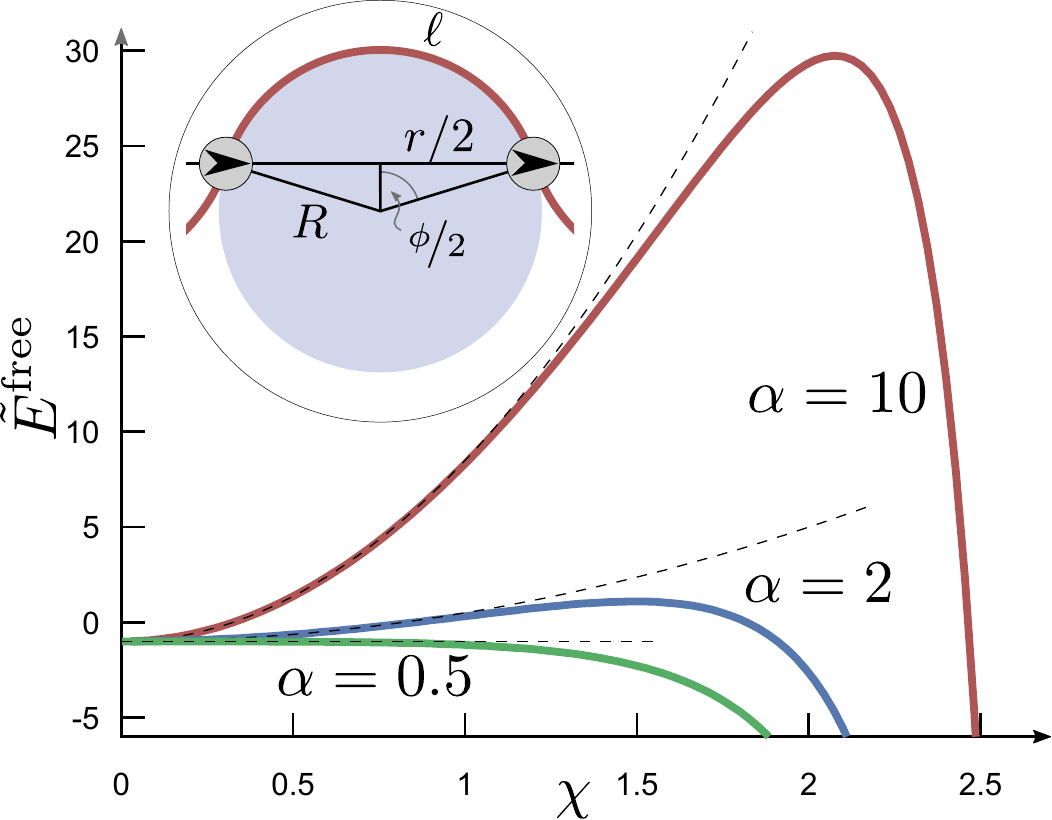}
\caption{Energy with magnetic moments {that are firmly attached but free to rotate} and a circularly bent rod, cf. eq. \eqref{eq:freeeb}, for various values of the rescaled bending stiffness $\alpha$. The corresponding dashed black lines show the approximation of eq. \eqref{eq:freeeb2}.\protect\linebreak Inset: Sketch of the corresponding geometry.}
\label{fig:ebfree}
\end{figure}

Let us first assume that the magnetic moments are attached in a manner that allows them to rotate freely, see inset in Fig. \ref{fig:ebfree}. In that case, the magnetic moments will align parallel to the distance vector $\vec{r}$. Thus, the magnetic energy in this case reduces to
{
\begin{align}
\frac{4\pi}{\mu_0} E_m^{\text{free}} &= - \frac{2 m^2}{r^3} = - \frac{2 m^2}{8 R^3 \sin^3 \frac{\ell}{2R}}
\end{align}}
where we employed the geometric relation $2R \sin \frac{\ell}{2R} = r$ to eliminate the distance $r$. As the arc-length $\ell$ of the filament is a parameter of the problem the radius $R$ is the only remaining variable and we are left with a one-dimensional optimization problem. The total energy is
$
E^{\text{free}} = E_m^{\text{free}}(R) + E_b(R)
$
with the bending energy $E_b(R)$ of the circular shape given by eq.\ \eqref{eq:ebcirc}.

We introduce the new variable $\chi=\frac{\ell}{2R}$ and rescale the energy by the absolute value\footnote{We use the absolute value because the energy of the straight configuration is negative and therefore interpreting the rescaled energy would become counter-intuitive otherwise.} of the energy of the straight, linear configuration $\frac{4\pi}{\mu_0}E^{\text{lin}}=-2m^2\ell^{-3}$ and find (see also Fig. \ref{fig:ebfree})
\begin{align}
\tilde{E}^{\text{free}} = \frac{E^{\text{free}}}{\left\lvert E^{\text{lin}}\right\rvert} &= - \frac{\chi^3}{\sin^3 \chi} + \alpha \chi^2  \label{eq:freeeb}
\end{align}
with the effective bending rigidity \begin{align}\alpha &= \varkappa \ell^2 \frac{4 \pi}{\mu_0} m^{-2}\text{.}\end{align} Inspecting the behavior for $\chi\sim \pi$ {one sees that the ``collapsed'' state will always be energetically the most favorable one as the energy is not bounded from below}.\footnote{This is an artifact of our model. In reality the $\chi\to \pi$- behavior will always be regularized, e.g. by steric interaction.} Whether this happens spontaneously, however, depends on the behavior at small $\chi$, i.e. whether or not there is an energy barrier to cross to get from the straight configuration to the collapsed one.  We expand $\tilde{E}^{\text{free}}$ for small $\chi$ to find
\begin{align}
\tilde{E}^{\text{free}} = - 1  + \left( - \frac{1}{2} + \alpha\right) \chi^2 + \mathcal{O}(\chi^3) \text{.} \label{eq:freeeb2}
\end{align}
Thus, we find that there is a ``renormalized'' stiffness due to the magnetic interaction $\tilde{\alpha} = \alpha - \frac{1}{2}$ which becomes {\em negative} for $\alpha < \alpha_c = \frac{1}{2}$. As the signs of both contributions to $\tilde{E}^{\text{free}}$ are unambiguous the sign of $\tilde{\alpha}$ determines the existence of an energy barrier. Thus, we deduce that $\alpha_c$ is the critical stiffness for the filament not to collapse.

Curiously enough, directly at this threshold the magnetic moments are effectively (to second order in $\chi$) bound by ``tether''-like interactions that restrict the maximal distance but do not energetically favor any distance within this maximal distance.

We note that the ``high density'' phase might be stabilized by higher order contributions to the elastic energy leading to a value of $\chi<\pi$. For example, we do not account for the elastic encapsulation of the magnetosomes or their finite size. However this does not alter the qualitative change of behavior as $\alpha$ crosses $\alpha_c$.

Plugging in biological values \cite{Kiani2015}, which in our notation can be given as {$\mu_0/(4\pi) m^2\sim 10^{-40} \,\mathrm{Nm^4}$, $\varkappa \sim 10^{-26}\, \mathrm{Nm^2}$}, we see that buckling is only relevant for filament lengths $\ell < \ell_c \approx 100\, \mathrm{nm}$ which is within the biologically relevant scale, thus indicating need for a stiff filament such as MamK. However, this is also notably close to the size of typical magnetosomes.
\section{Fixed alignment} \label{sec:fixed}
As a variation of the problem, we consider the situation in which the magnetic moments are fixed to be tangential to the filament. This situation leads to a torque exerted by the magnetic moments onto the rod. We refer to this torque as ``attachment torque'' to distinguish it from the torque exerted by the rod upon change of curvature. The quantitative value of this torque depends on microscopic details of the attachment (such as distance to the central fiber). As we are more interested in a generic picture we only study the two limiting cases of very low, negligible attachment torque, where the magnetic moments simply align along the tangents of the circular configuration studied before, and of very high, dominant attachment torque, where the magnetic moments force the tangents of the rod to be parallel thus enforcing a new class of configurations which we call the loop shape.
 
\subsection{Negligible attachment torque} \label{sec:fixedlow}

\begin{figure}
\includegraphics[width=.975\linewidth]{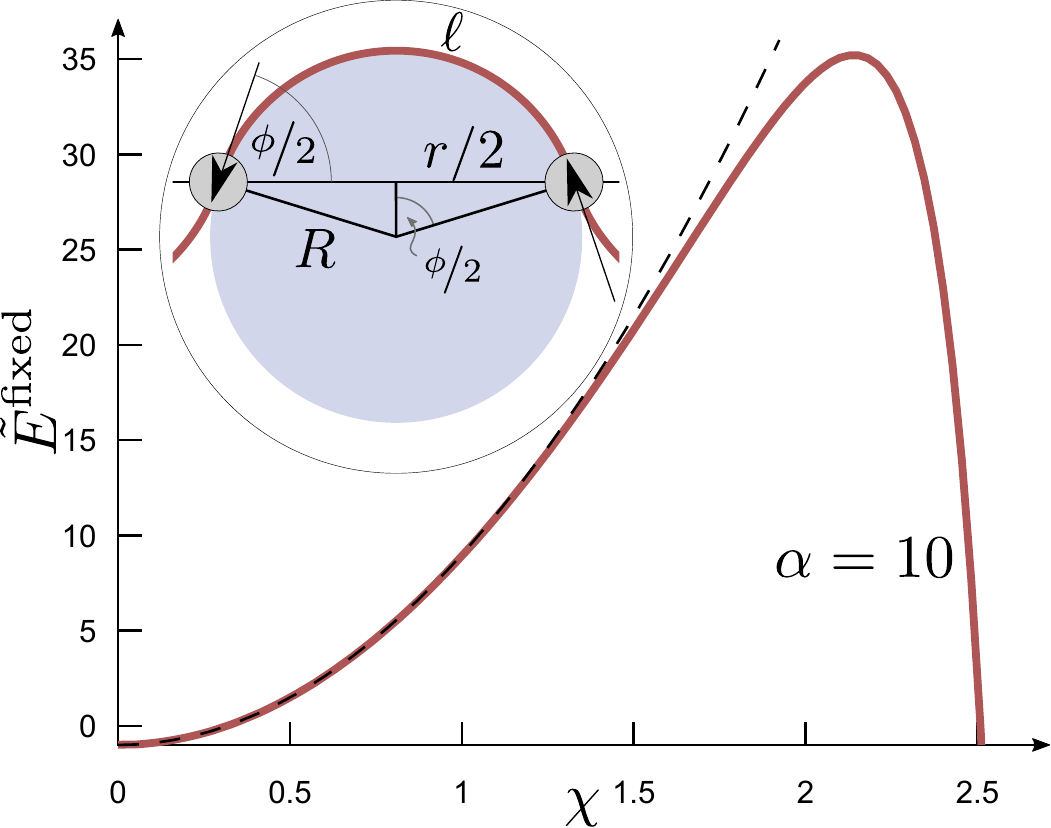}
\caption{Energy with magnetic moments exerting a negligible attachment torque and a circularly bent rod, cf. eq. \eqref{eq:fixedeb}. The dashed black line shows the approximation to quadratic order.\protect\linebreak Inset: Sketch of the corresponding geometry.}
\label{fig:ebfixed}
\end{figure}

First, we study the case of negligible attachment torque which leads to the situation depicted in the inset of Fig.\ \ref{fig:ebfixed}. Proceeding in completely analogous fashion to the case before and employing suitable geometric relations, we find for the rescaled energy in this case{
\begin{align}
\tilde{E}^{\text{fixed}} &= - \frac{\left(\cos^2 \chi+1\right) \chi^3}{2\sin ^3 \chi} + \alpha \chi^2 \label{eq:fixedeb}
\end{align}
which is plotted in Fig.\ \ref{fig:ebfixed}. This energy expands to
$\tilde{E}^{\text{fixed}} = -1 + \alpha \chi^2 + \mathcal{O}(\chi^3)$ 
for small $\chi$.
Thus, in this case the bending rigidity is {\em unchanged}: there is always an energy barrier preventing the spontaneous collapse. However, it is notable that this also means that there is {\em no additional stiffness} due to the magnetic interactions.}

\begin{figure}
\includegraphics[width=.975\linewidth]{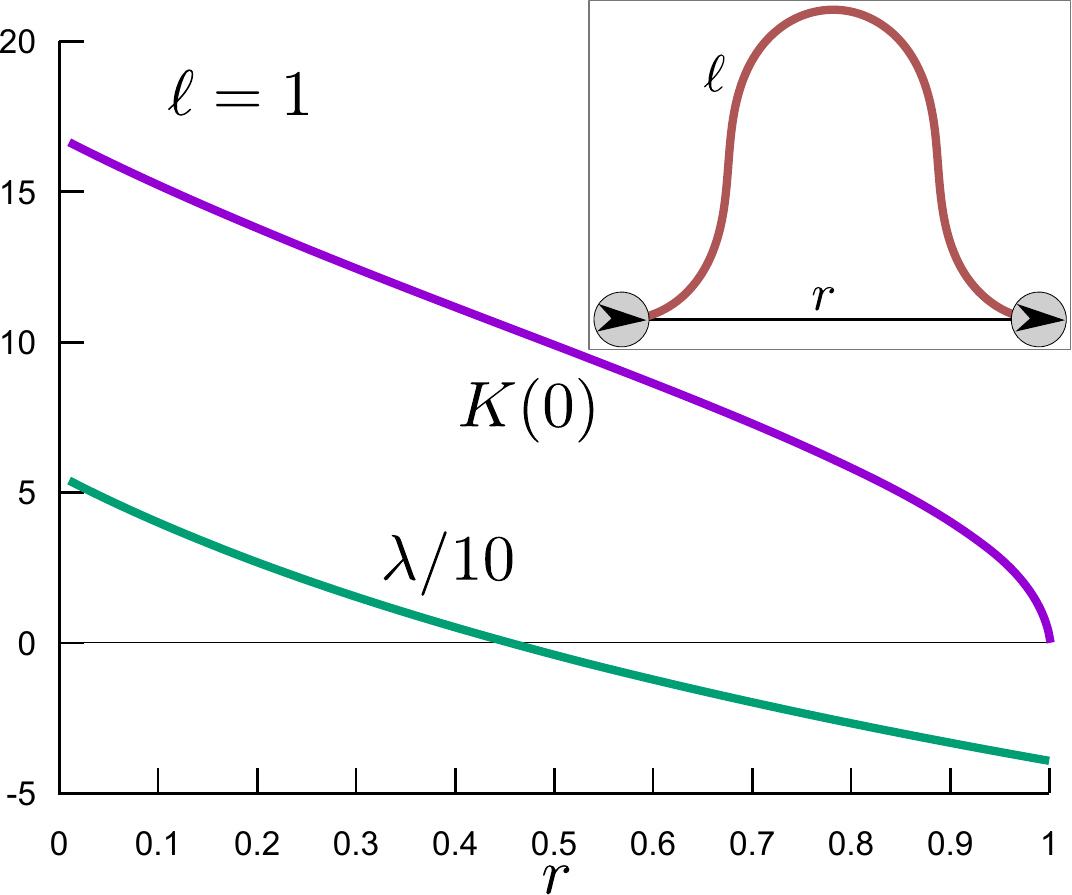}
\caption{Initial condition $K(0)$ and Lagrange parameter $\lambda$ (rescaled for optical purposes) of the loop-solution as a function of the final distance $r$. {This solution branch is relevant if a high attachment torque is exerted onto the filament forcing it to be parallel to the distance vector between the magnetic moments at the two ends.} The actual shape of the rod is then easily integrated from eq.\ \eqref{eq:dgl} (which is closed by giving $K(0)$,$\lambda$ and $\partial_s K(0)=0$) and the Frenet-Serret-formulae \eqref{eq:fs}.\protect\linebreak Inset: {Exaggerated visualization} of the loop shape. Details given in text.}
\label{fig:loopparams}
\end{figure}

The two subproblems of free alignment and fixed alignment with negligible attachment torque coincide when {considering undulations with wavelengths spreading of a larger number of the magnetic moments}\cite{Vella2013,Kiani2015} as the optimal alignment turns out to be tangential to the curve of the filament. Then, the properly rescaled energy is (for $N\gg 1$) given by
$
\tilde E^{N\gg 1} = \frac{\alpha \chi^2}{N^2} - \frac{\chi^3 (\cos^2 \chi+1)}{2 \sin ^3 \chi}
$
if one continues to only take next-neighbor interactions into account.  As one would expect, the relative contribution of the bending term is smaller when considering buckling patterns that spread over multiple magnetic moments. However, it turns out that the change in magnetic energy due to this is non-negative for the onset of buckling (small $\chi$); contrary to what would be needed to lead to spontaneous buckling.

This problem has been studied with more mathematical rigor using the full interaction \cite{Vella2013,Kiani2015} with the result that there is a {\em positive} contribution to the bending stiffness, which qualitatively is in line with the rough calculation presented here. Thus, it is never energetically advantageous to buckle the magnetic moments out of the axis given by their magnetization. This strongly justifies our limitation to the problem of two magnetic moments as the relevant buckling here is the buckling of the connective rod between to magnetic moments and undulations on larger lengthscales (with slightly tilted magnetic dipoles) will lead to an additional magnetic bending rigidity.

\subsection{High attachment torque} \label{sec:fixedhigh}

The other prominent limiting case is the one of high attachment torque leading to parallel magnetic moments but also parallel tangents at the two ends of the rod. 

This way, effectively there is again no torque acting onto the rod, as the {tangents to the rod} and the magnetic moments are parallel. Additionally, we demand periodicity, $K(s=0)=\pm K(s=\ell)$, $\partial_s K(s=0)=\partial_s K(s=\ell)=0$.

The relevant (i.e. energetically lowest) solution branch is exemplified by the curve shown in the inset of Fig. \ref{fig:loopparams}, which we refer to as the loop solution. It is numerically straightforward to follow this branch (see Fig.\ \ref{fig:loopparams}) for any $r \in [0,1]$ and compute the energy. However, we can also make analytical progress with some reasonable approximations.

\begin{figure}
\includegraphics[width=8cm]{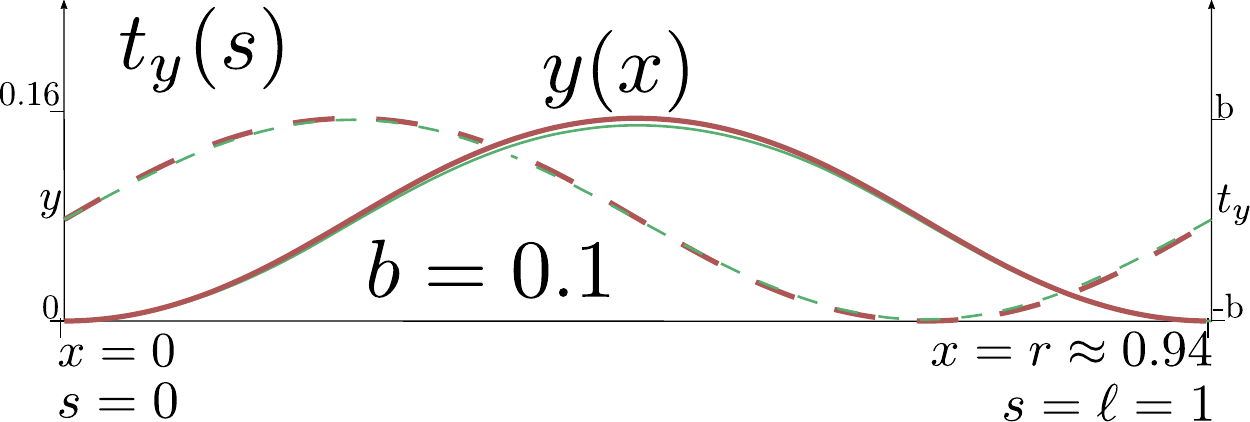}
\caption{{Comparison} of the approximation (thin green lines) of eq.\ \eqref{eq:approx} (for $b=0.1$) with an actual numerical solution (thick red lines). Shown are (using arbitrary scales) the shape $y(x)$ (solid lines) and the $y$-component of the tangent vector $t_y(s)$ as a function of the arc-length (dashed lines). {For a compact presentation, the two different data sets use different axes (cp. labeling in the figure).} The closer the end-to-end distance $r$ gets to the maximal value of $\ell$ the better the approximation of eq.\ \eqref{eq:approx} becomes. This gives confidence that the small $b$-behavior (as outlined in the text) captures the essence of the onset of buckling. }
\label{fig:approx}
\end{figure}

\begin{figure}
\includegraphics[width=.975\linewidth]{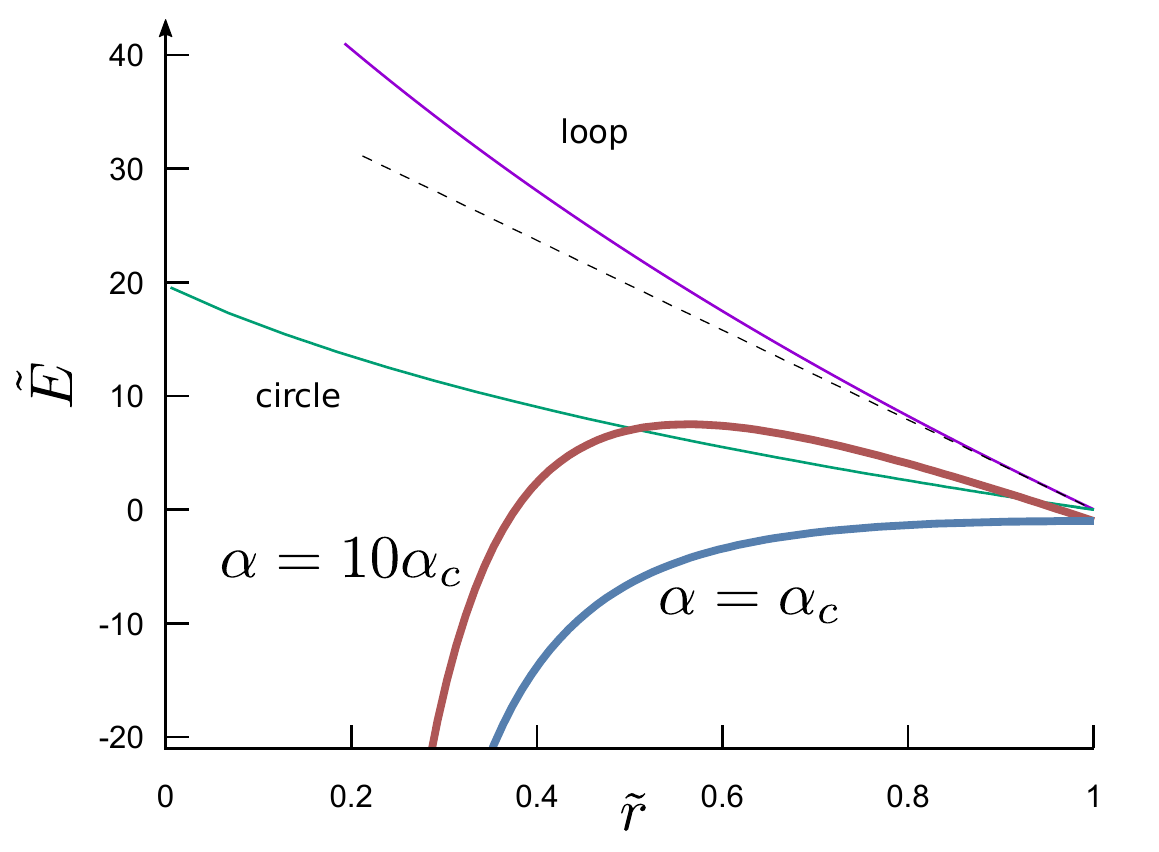}
\caption{Energy with fixed magnetic moments and a loop-like bent rod. The thin solid lines show the bending energy (rescaled for clarity) in the loop configuration and (for comparison) in a circle. We find that to high accuracy the relative difference is given by a factor of $\pi$. We are, however, at the moment not able to give an analytical justification for this. The dashed black line shows the approximation of eq. \eqref{eq:ebloop2}. The thicker solid lines show the total approximated energy, cp. eq. \eqref{eq:ebloop}.}
\label{fig:ebloop}
\end{figure}

Inspecting the geometry of the solutions, see Fig.\ \ref{fig:approx} and using a frame of reference in which the rod is within the $xy$-plane and $t_x(0)=1$,  we find that a fairly accurate description is given by approximating the $y$-component of the tangent by 
\begin{align}
t_y(s) &= b \sin \left(\frac{2 \pi}{\ell} s\right) \text{.} \label{eq:approx}
\end{align}
Here $b$ is a parameter giving a family of shapes. Integrating the elastica equation leads to a relation for the distance $r(\ell)$ as a function of $b$, see below. From the length constraint $\lvert \vec t\vert =1$ we conclude
$
t_x(s) = \sqrt{1-b^2 \sin^2 \left(\frac{2\pi}{\ell}s\right) } \text{.}
$
{Using this and the $y$-component of the first Frenet-Serret formula \eqref{eq:fs} leads to
\begin{align}
\frac{2\pi b}{\ell} \cos \frac{2\pi s}{\ell} &= K \sqrt{1-b^2\sin^2 \frac{2\pi s}{\ell}}
\end{align}
or}
\begin{align}
K^2 &=  \frac{4\pi^2\frac{b^2}{\ell^2}\cos^2 \frac{2\pi s}{\ell}}{1-b^2\sin^2 \frac{2\pi s}{\ell}}= \frac{2\pi^2 b^2}{\ell^2} \left( \cos \frac{4\pi s}{\ell} +1 \right) +\mathcal{O}(b^4) \text{.} 
\end{align}
From this, we find the corresponding bending energy\footnote{We use $\protect\langle \cdot \protect\rangle$ for averages over the contour of the filament}
\begin{align}
E_b^\text{loop} &= \frac{\varkappa}{2} \ell \langle K^2 \rangle= \varkappa \frac{b^2 \pi^2}{\ell}+\mathcal{O}(b^4) \text{.} \label{eq:Eb} 
\end{align}


{Moving on, we can integrate the tangent to find the distance $r=\int_0^\ell\!\mathrm{d}s\,t_x(s)$ as a function of the parameter $b$ in terms of the complete elliptic integral $\mathcal{E}(m)=\int_0^\frac{\pi}{2} \!\mathrm{d}x\, \sqrt{1-m\sin^2 x}$
\begin{align}
r&= \frac{\ell}{2\pi} \int_0^{2\pi} \!\mathrm{d}q \sqrt{1-b^2\sin^2 q} = \frac{\ell}{\pi} \int_0^\pi \!\mathrm{d}q \sqrt{1-b^2\sin^2 q}  \\
&= \frac{\ell}{\pi} \left( \mathcal{E}(b^2) + \int_{\frac{\pi}{2}}^{\pi}\!\mathrm{d}q \sqrt{1-b^2 \sin^2 q} \right)\\
&=  \frac{\ell}{\pi} \left( \mathcal{E}(b^2) + \int_{0}^{\pi/2}\!\mathrm{d}p \sqrt{1-b^2}\sqrt{1-\frac{b^2}{b^2-1} \sin^2 p} \right)\\
&=\frac{\ell \mathcal{E}(b^2)}{\pi}{+}\sqrt{1{-}b^2}\mathcal{E}(\frac{b^2}{b^2{-}1})= \ell {-} \frac{b^2 \ell}{4} {+} \mathcal{O}(b^4) \text{.} \label{eq:r}
\end{align}}
Combining eqs. \eqref{eq:Eb} and \eqref{eq:r} we can eliminate $b$ to find
\begin{align}
E_b^\text{loop}(r) &= 4\pi^2\varkappa\frac{\ell-r}{\ell^2} + \mathcal{O}((l-r)^2)  \label{eq:ebloop2}
\end{align}
and thus the full {rescaled} energy can be approximated by $\tilde{E}^\text{loop}(r) \approx -\frac{\ell^3}{r^3}+4\pi^2\frac{\varkappa\ell^2}{2 m^2} \frac{\ell}{\ell} \frac{\ell -r}{\ell}$ or
\begin{align}
\tilde{E}^\text{loop}(\tilde r) &= -\tilde{r}^{-3} +2 \pi^2 \alpha (1-\tilde{r}) + \mathcal{O}( (1-\tilde r)^2) \label{eq:ebloop}
\end{align}
with $\tilde r= r/\ell$. Inspecting the behavior at $\tilde r\approx 1$ we find again a critical stiffness $\alpha_c = \frac{3}{2\pi^2} \approx 1/6$. Thus, the resistance to buckling is noticeably (and somewhat expectably) higher than before in the case of free alignment. We show the energies in Fig. \ref{fig:ebloop}. The energy to all orders of $1-\tilde{r}$ could be computed by adding the magnetic contribution to the numerically found values of the bending energy. However, we are mostly interested in the onset of buckling and therefore this analysis suffices.

The result derived here, should be considered as an upper bound to an $\alpha_c$ expected for any attachment mechanisms that exert some kind of torque onto the rod. It is of the same order of magnitude as the result we obtained for free alignment with $\alpha_c = \mathcal{O}(1)$. Thus, our estimates of biological relevant scales from before hold independently of the intricacies of the attachment.

\section{Conclusion}

In this work we have studied the buckling behavior of a stiff, elastic rod due to the compressive forces exerted by magnetic moments that are placed along its contour. This is a (highly idealized) model of the magnetosome chain within  magnetotactic bacteria. As we were mostly focusing on an analytical treatment we neglected thermal fluctuations (which should be justified on lengthscales below the rod's persistence length) {as well as any possible long-wavelength deformations consisting of rotated magnetic moments} and reduced the problem to a two-dimensional one. {Also, we limited our analysis to the case of permanent magnetic moments (as is biologically relevant). The more intricate features of magnetosomes with paramagnetic behavior is left to future studies}.

{We discussed various scenarios of the nature of attachment of these magnetic moments and found analytical expressions for the critical stiffness of the rod needed to sustain a straight configuration. In the simplest case, that the magnetic moments are attached in a way that they can exert forces onto the rod but no torques. In that case we find a critical bending strength $\kappa_c \sim \mu_0 m^2 /\ell$. By contrast, if the magnetic moments align parallel to the curve of the rod without exerting a significant torque onto it, we find that there is no spontaneous buckling, irrespective of the strength of the magnetic interactions. Finally, we considered the other limit in which the torque from the magnetic moments is always dominant. We studied this case by numerically solving the Euler elastica equations and  find qualitatively the same behavior as in the first case (with a reduced critical bending strength, as this requires more heavily bent solutions).}

{
A biological system that realizes such elastic rods with magnetic moments, are the magnetosome chains of magnetotactic bacteria, which consist of membrane-enclosed magnetic nanoparticles attached to a filament of the bacterial actin-like protein, MamK. It is difficult to say which of the scenarios we discussed provides the best description of this system (in addition, the biomechanical mechanisms might to some extent be species-dependent), however, we expect the existence of a buckling transition to be the generic case. Such a transition is then governed rather universally by critical parameters following the expression derived here, up to minor differences in prefactors. }

From an estimate of scales we see that buckling of the connective filament is within reach of the biological values, thus indicating the functional relevance of the bending stiffness{, which stabilizes the filament against buckling, thus providing the ability to spread the magnetic moments to create a well-aligned structure.
We expect this to be particularly relevant in (biologically transient) developmental stages where the typical distances between magnetosomes can be relatively large, as  they are not as densely packed as in the mature magnetosome chain and where chains may exhibit large gaps \cite{Faivre2010}. }

\section{Authors contributions}
All the authors were involved in the preparation of the manuscript.
All the authors have read and approved the final manuscript.

\appendix
\section{Appendix: Derivation of the elastica equation}
We give a brief derivation of \eqref{eq:dgl}, mostly following ref.~\citenum{Singer2008}. We consider the effective energy functional
\begin{align}
E_V=\frac{\kappa}{2} \int_0^L\!\mathrm{d}s\, \left[ \left(\partial^2_s \vec q\right)^2 + \eta(s) \left(( \partial_s \vec q)^2 -1 \right) \right] 
\end{align}
wherein the second term corresponds to the {\em local} inextensibility constraint and the multiplier function $\eta(s)$ will be determined later on. We now demand that $E_V$ is stationary under a variation $\vec q\to \vec q + \epsilon \vec v$ (with $\vec v$ vanishing at the ends) of the space-curve $\vec q$ which leads (after integration by parts) to the Euler-Lagrange equation
\begin{align}
0&=\partial^4_s \vec{q} - \left(\partial_s \eta(s)) \partial_s \vec q) + \eta(s) \partial_s^2 \vec q\right) \text{.}
\end{align}
From the Frenet-Serret-formulae, eq. \eqref{eq:fs}, we know that
\begin{align}
\partial_s \vec q &= \vec t\\
\partial^2_s \vec q &= K \vec n\\
\partial^3_s \vec q &= (\partial_s K) \vec n - K^2 \vec t\\
\partial^4_s \vec q &= (\partial_s^2 K) \vec n - 3 K (\partial_s K) \vec t + K^3 \vec n 
\end{align}
and we can use this to write the Euler-Lagrange equation as
\begin{align}
0&=(\partial^2_s K - \eta K + K^3)\vec n + ( -3K(\partial_s K) - \partial_s \eta)\vec t \text{.}
\end{align}
Both terms have to vanish individually as $\vec n \perp \vec t$. From the second term, we gather $\eta(s) = -\frac{3}{2} K^2 + \lambda$ with some {\em constant} $\lambda$. Plugging this into the first term, we find
\begin{align}
0 = \partial_s^2 K + \frac{1}{2} K^3 - \lambda K \text{,}
\end{align}
which is the equation given in the main text.
\end{document}